\documentclass[twocolumn,showpacs,preprintnumbers,amsmath,amssymb]{revtex4}
\usepackage{tabularx,graphicx}

\usepackage{color}
\usepackage{hyperref}
\hypersetup{
    colorlinks=true,
    linkcolor=blue,
    filecolor=blue,      
    urlcolor=blue,
}

\usepackage{color,soul}

\usepackage{ulem}   

\begin{document}

\newcommand{\beq}{\begin{equation}}
\newcommand{\eeq}{\end{equation}}
\newcommand{\beqn}{\begin{eqnarray}}
\newcommand{\eeqn}{\end{eqnarray}}
\newcommand{\bmath}{\begin{subequations}}
\newcommand{\emath}{\end{subequations}}
\newcommand{\bra}[1]{\langle #1|}
\newcommand{\ket}[1]{|#1\rangle}

\title{Reply to ``Comment on `Nonstandard  superconductivity or no superconductivity in hydrides under high pressure' ''}

\author{J. E. Hirsch$^{a}$  and F. Marsiglio$^{b}$ }
\address{$^{a}$Department of Physics, University of California, San Diego,
La Jolla, CA 92093-0319\\
$^{b}$Department of Physics, University of Alberta, Edmonton,
Alberta, Canada T6G 2E1}

\begin{abstract} 
In Ref. \cite{nonstandard} we surveyed the known hydride superconductors, and compared their resistive behavior to that of typical
known superconductors, including conventional (e.g. NbN and MgB$_2$) and unconventional (e.g. YBCO) superconductors, and concluded that the behavior
of the hydrides was indicative of nonstandard or no superconductivity. In the preceding
comment, Talantsev, Minkov, Balakirev and Eremets \cite{talantsev_comment} (arXiv:2311.07865) claim that we presented a ``{\it flawed analysis and a selective and inaccurate report
of published data}.'' Here we show that this claim is wrong.
\end{abstract}
\pacs{}
\maketitle 

\section{Introduction}
In their Comment, Talantsev, Minkov, Balakirev and Eremets \cite{talantsev_comment} claim (first sentence in their abstract, and elsewhere in the
Comment) that
in our paper  \cite{nonstandard} we
{\it ``assert that hydrogen-rich compounds do not exhibit superconductivity''}. That claim is non-factual: our paper claimed instead that
{\it ``Our
results indicate that either these materials are unconventional superconductors of a novel kind, which we term
“nonstandard superconductors,” or alternatively, that they are not superconductors''}, and that
{\it ``If the second is the
case, which we believe is more likely, we suggest that the signals interpreted as superconductivity are either
experimental artifacts or they signal other interesting physics but not superconductivity.''}
We will assume in what follows that their statement that our claim 
 ``{\it relies on a flawed analysis and a selective and inaccurate report
of published data''} refers to our true claim, not an imagined one. To refute that claim, we follow the sequence of their sections II-VI, their titles given in italics in the listing below, and discuss their figures which purport to support their claim.\\

\section{detailed analysis}

1) {\it The transition width definition.}\\

Our observation of nonstandard behavior in the hydrides compared to that in known superconductors is based on observation of {\it qualitative}
trends. That such an approach is required is exemplified by the myriad examples of resistivity vs. temperature curves with knees, tails, kinks, non-monotonic behavior, etc, for which an arbitrary rigid definition of width such as given in Eq. (1) of the Comment is not  appropriate. That is,
choosing $\Delta T_{0.1-0.9}$ as in Eq. (1) of the Comment, versus $\Delta T_{0.15-0.85}$ or  $\Delta T_{0.05-0.95}$, would yield qualitatively different answers. For all the examples discussed in our paper \cite{nonstandard}, both standard superconductors and hydride materials, the curves of
resistance versus temperature for varying magnetic fields
from which we extracted the widths shown in our Fig. 11 are shown in our paper also, enabling readers to easily perform their own analysis
with their favorite  definition of transition width such as Eq. (1) in the Comment, and compare with our estimates given in Fig. 11 of \cite{nonstandard}. Surprisingly, this
was not done in the Comment.\\


2) {\it Type I superconductors.}\\

We do not believe that the hydrides are type I superconductors, nor does anyone else as far as we know, so we consider this discussion in the Comment
to be irrelevant. Note that if these hydrides really were type I superconductors with such high $T_c$'s, 
that don't expel magnetic fields and trap magnetic fields according to what is reported by the authors of the Comment in other
publications \cite{e2021p,etrappedp}, then indeed they would be $very$ nonstandard superconductors.\\

3) {\it Type II superconductors (non-hydrides)}\\

Various examples of resistivity widths for non-hydride type II superconductors have been provided in the Comment. Shown in their Fig. (3a) are results for $2.9$ nm thin films of MgB$_2$, with fields applied both
perpendicular and parallel to the plane of the thin film, reported in Ref. \cite{mgb2thin}. There is an initial decrease of transition width as a function of applied field, followed by the standard increase with magnetic
field. 
These films have thickness of only a few unit cells, smaller than the superconducting coherence length and much smaller than the London penetration depth. Clearly the anomalous initial decrease of transition width with 
magnetic field highlighted in the Comment is a property connected to ultrathin films. The hydride samples have
thicknesses that are larger by three orders of magnitude and surely resemble the polycrystalline samples used in
the study referred to in our paper and many other such studies  more than they resemble highly oriented ultrathin films.

Fig. (3b)  in the Comment are results reported in Ref. \cite{wang2010} for polycrystalline MgB$_2$ samples doped by SiC, that also show an upturn for low fields.
The authors of Ref. \cite{wang2010}  discuss this anomalous feature of their data in great detail, and conclude that
 {\it `` the origin of the low field $\Delta T$  upturn is due to multiple superconducting
transitions resulting from the replacement of B and/or Mg atoms by other elements.''}
 We cannot rule out that multiple transitions in the hydrides could also conceivably give rise to such an upturn. 
 We are not aware however of such upturns in the hydrides, instead our paper  Ref. \cite{nonstandard} highlighted the fact that in 
 many cases the transition widths are approximately  constant as function of field.

\begin{figure} []
\resizebox{8.5cm}{!}{\includegraphics[width=6cm]{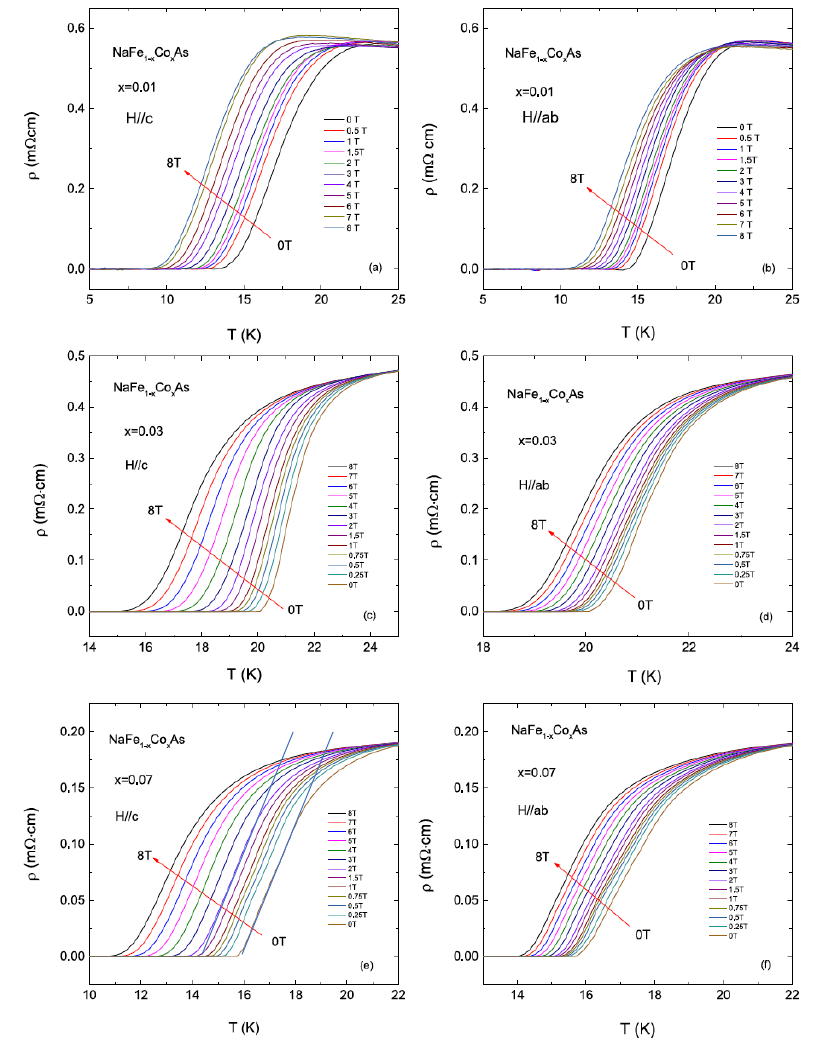}} 
 \caption {Fig. 4 from Ref. \cite{choi2017}. Note the variation of the temperature scales for panels (a) and (b) ($x=0.01$), compared to (c) and (d) ($x=0.03$) and to (e) and (f) ($x=0.07$).
 At $x=0.01$ the material has entered an antiferromagnetic state, so application of a magnetic field also affects the underlying magnetic state.
 }
 \label{reply_figure1}
 \end{figure}

In Fig. (3c) in the Comment, a figure is presented for a pnictide, NaFe$_{1-x}$Co$_x$As. In this case, referring to the Comment's Ref. 27, which we cite as Ref. \cite{choi2017}, the authors ``cherry-picked''  one composition ($x = 0.01$) for an applied field parallel to the c-axis, from Fig. (4a) of Ref. \cite{choi2017}. However, as subsequent panels in that
same Fig. 4 of Ref. \cite{choi2017} show, reproduced here in Fig. (\ref{reply_figure1}), all higher compositions ($x=0.03$ and $x=0.07$) 
clearly show the normal broadening as a function of applied field. 
 This compound at this stoichiometry is a poor choice, because, as the 
authors of  \cite{choi2017} indicate, the compound with $x=0.01$ is actually also antiferromagnetic, with a N\'eel temperature of 33 K, not too far above the superconducting transition
temperature. The authors also point out that the superconducting transition at $x=0.01$ is extremely broad for this reason (compare with panels (c) and (e) in Fig. (\ref{reply_figure1})), so
one can imagine that a reduction in the width with applied field could well reflect the impact of that field on the underlying antiferromagnetic state. 
  Since no antiferromagnetic transitions are
expected in the hydride materials this is hardly a suitable
choice of material with which to critique our results, especially leaving out the higher dopant values.

\begin{figure} []
\resizebox{8.5cm}{!}{\includegraphics[width=6cm]{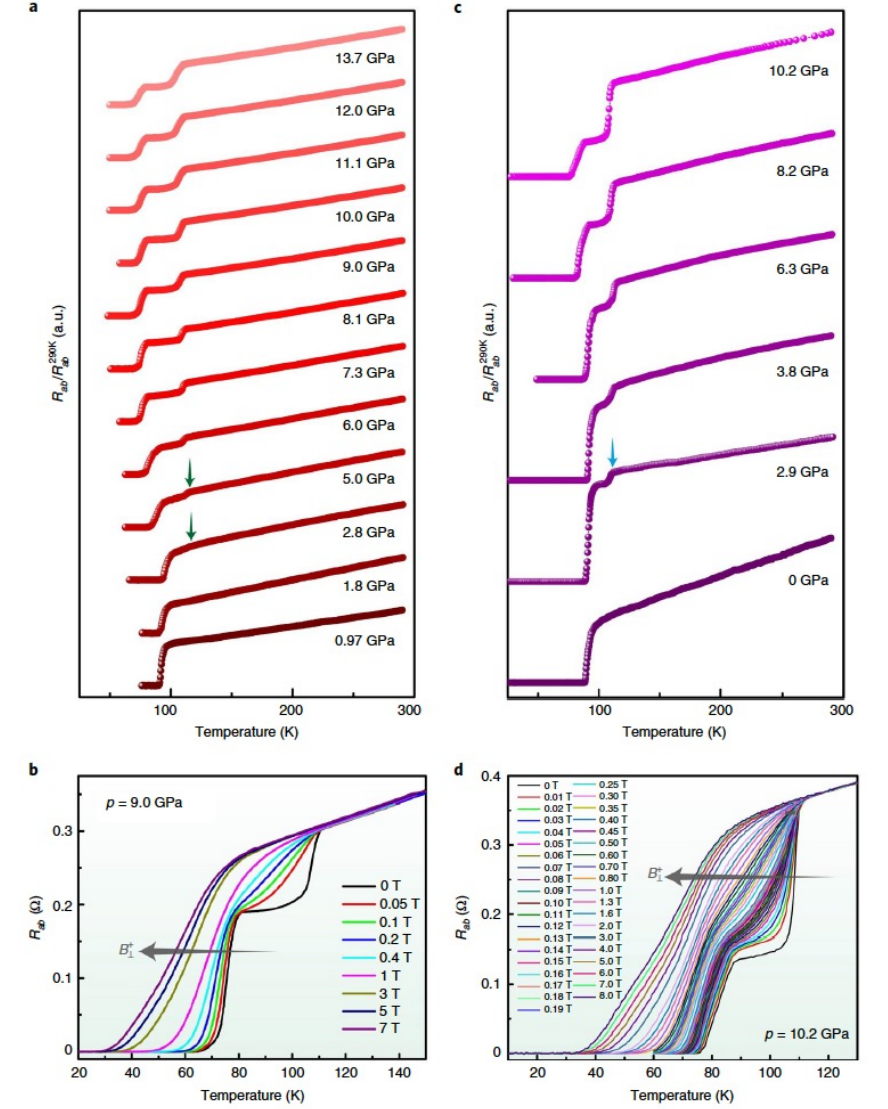}} 
 \caption {Fig. 1 from Ref. \cite{guo2020}.  It is a strikingly poor material with which to
 conclude anything about transition widths, particularly at low applied field where a pronounced knee is present in the resistivity data (panels (b) and (d)). Note the clear indication
 of two transitions at 9 GPa (in panel (a)) so the comments from sintered MgB$_2$ paper (Wang et al. \cite{wang2010}) about the ``abnormal'' decrease in transition width could
 well apply.}
 \label{reply_figure2}
 \end{figure}

In Fig. (3d) in the Comment, an example of a cuprate material, Bi$_2$Sr$_2$CaCu$_2$O$_8$, with applied pressure of $9$ GPa is given, using data 
obtained from the Comment's Ref. (28), cited here as Ref. \cite{guo2020}.
Again, this is a poor choice of material with which to critique our work, as it was used by the authors of Ref. \cite{guo2020} to observe ``a pressure-induced crossover
from two- to three-dimensional (2D to 2D) superconducting states,'' resulting in ``Berezinskii-Kosterlitz-Thouless-like behavior''. Therefore a significant amount of nonstandard physics
is occurring in this material, that also exhibits multiple transitions. We show Fig.~(1) from that paper here as Fig.~(\ref{reply_figure2}). The displayed resistance versus temperature curves clearly indicate
abnormal behavior at low applied fields, along with the expected broadening with field at higher magnetic fields, where the resistivity curves lack the abnormality present at lower fields.\\

4) {\it Highly compressed scandium.}\\

These measurements are at high pressure, and only under the highest pressure measured (262 GPa) is there an indication from Fig.~(4) in the Comment that the width decreases
with increasing fields at very low fields. But this is precisely where a significant knee occurs (see Fig. 8d in the supplementary material of the Comment), and so concerns about the
interpretation of this initial decrease echo those above.\\

5) {\it Hydride superconductors.}\\

\begin{figure} []
\resizebox{8.5cm}{!}{\includegraphics[width=6cm]{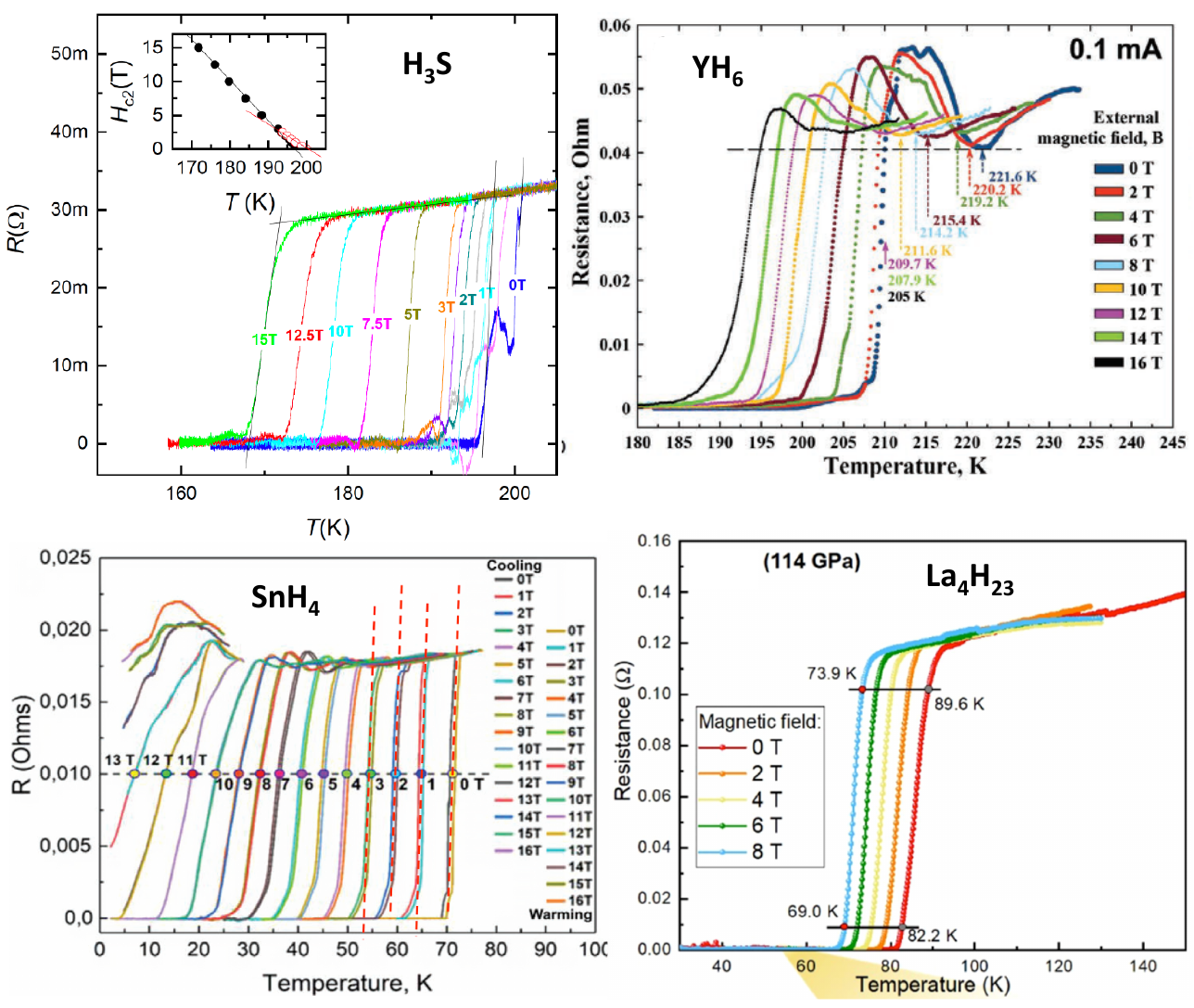}} 
 \caption {Some of the resistance versus temperature profiles used in Fig. 5 of the Comment, and one additional figure (lower right panel). A detailed description is given in the text. Briefly, all of these results show
 anomalous behavior over some temperature and/or magnetic field range, and determining a transition width using Eq.~(1) of Ref. \cite{talantsev_comment} is clearly problematic.
 In the lower left panel we have included (red) {\it parallel} lines to indicate that the widths of these transitions are not changing at low applied magnetic field.
In the lower right panel we show recent results for $LaH_{23}$ from Ref. \cite{recent}.}
 \label{reply_figure3}. 
 \end{figure}

Figure~5 of the Comment shows  resistance width versus applied field constructed by the authors based on their Eq.~(1), but  applied to very complicated resistance versus temperature measurements. It is not clear what the purpose of this Figure is. In Fig.~\ref{reply_figure3} we show a number of resistance versus temperature profiles used by the authors of the Comment in their Fig. 5.
The upper left panel shows the data from Mozaffari et al. \cite{mozaffari2019} on H$_3$S for a variety of applied magnetic fields
whose width is shown in Fig. 5a of the Comment. It is evident
that there are difficulties with interpretation, as a significant kink exists in the low field (including zero field) data. Mozaffari et al. \cite{mozaffari2019} attribute this to inhomogeneities, but do
not have an explanation for why application of a sufficiently high magnetic field seemingly removes these inhomogeneities, or why there was a time dependence for these resistance
measurements. Whatever the explanation, at the very least, their resistance versus temperature measurements constitute strong evidence for nonstandard superconductivity and,
in our opinion, it is inprudent to try to analyze the transition width as a function of applied magnetic field for such data.

To construct the result in their Fig.~(5b), Talantsev et al. use  resistance versus temperature data from Troyan et al. \cite{troyan2021} for YH$_6$. We show this data
in the upper right panel of Fig.~(\ref{reply_figure3}); this figure clearly shows anomalous characteristics.
So once again, if we cannot qualitatively understand the resistance versus temperature measurements, it seems ill-advised to apply a formula such as 
the Comment's Eq.~(1) to analyse their widths.

 In panel (c) of Fig.~5 of the Comment, a number of compounds are included. The (La,Ce)H$_9$ data (red diamonds) are very similar to the ones we presented
 for LaH$_{10}$ (black squares) and therefore support our claim of very little or decreasing broadening with applied field. The plotted results for SnH$_4$ are again
 misleading as a perusal of the resistance versus temperature data indicate. These data, from Troyan et al. \cite{troyan2023}, are included in the lower left panel of our Fig.~\ref{reply_figure3},
 together with four {\it parallel} red lines that represent the slopes of the resistance drops for the three lowest magnetic fields quite well. The fact that they are {\it parallel} indicates
 that the width is not changing at all with increasing field, contrary to the claim made in the Comment, where an  increasing slope is indicated as a function of magnetic field.
 As the field increases further the resistance drops begin to broaden, but largely because of an increasing tail at the low temperature end. This is most apparent for very high applied
 fields, where the normal state resistance takes a semiconductor-like upward turn and eventually the near-zero resistance limit is never achieved. To attribute this abnormal behavior simply
 to transition width broadening is very misleading. Finally, the HfH$_x$ results shown in panel (c) of Fig.~5 of the Comment show both a very small width
 for zero field and a very small increase as a function of
 applied magnetic field, for fields as large as   $B_{appl}/B_{c2}=0.3$. This is qualitatively different from the broadenings seen in standard superconductors shown in 
 Fig. 11a of our paper Ref. \cite{nonstandard}, hence we consider these results to be consistent with the message in our paper.
 
 To conclude, we show in the lower right panel of Fig. 3   recent results for $La_4H_{23}$ from Ref. \cite{recent}, showing
 anomalous narrowing of the resistance drop with increasing field.
 The authors of Ref. \cite{recent}  comment that {\it ``The sample demonstrates the absence (and
even negative value) of broadening of superconducting transitions in a magnetic field, as previously
observed for yttrium ($YH_6$) and lanthanum-yttrium ($(La,Y)H_{10}$) hydrides''}, confirming the theme of our paper Ref. \cite{nonstandard} that
 the anomalous behavior in resistance width that we pointed out is common in so-called hydride superconductors.
  \\
 
\section{Summary and conclusion}
 
To conclude this Reply, we address each point of the {\it ``Detailed explanation of the primary message of the Comment''} in its Sect. I, where the authors claim to
{\it ``demonstrate the following''}:

{\it ``1. There is no universal in-field transition width  dependence that can serve as the
criterion for determining the existence of the superconductivity in any given material''}


 There is a near-universal broadening of the resistive transition versus field in standard type II superconductors,
whose origin is well understood \cite{nonstandard,blatter}. As shown in this Reply, for all examples of ``anomalous'' behavior for standard superconductors presented in the Comment
we have shown that there are clear reasons for the anomalous behavior that are not expected to be present in the hydrides.
Absence of such broadening in many hydride materials suggests that the resistance drops seen in hydrides are not due to superconductivity
but due to other reasons \cite{hmnature,dc,enormous,vdp}.

{\it ``2. Hydrides at high pressures exhibit a similar in-field transition width  dependence to
other superconductors, such as MgB2''}

The typical behavior of $MgB_2$ samples is very standard, the few samples that show anomalous behavior do it for reasons that don't apply
to the hydrides.

{\it ``3. The authors employed undisclosed definitions/methods for extracting:
 the transition temperature and  the transition width from raw experimental resistance data''}

We showed figures for the resistance curves for all cases discussed in our paper, the qualitative behavior that we highlighted is very clear from those figures.
The authors of the Comment have not shown that their criteria for transition width gives qualitatively different behavior than what we showed in
Fig. 11 of our paper Ref. \cite{nonstandard}.

{\it ``4. The authors[1] selectively presented datasets that support their claims while excluding datasets
that contradict them''}

We presented a representative set of examples available at that time. Since then other materials have been measured, and several of them
show nonstandard behavior as discussed in the previous section.

{\it ``5. The authors[1] misinterpreted the magnetic phase diagrams of superconductors by overlooking
the existence of the intermediate state in type-I superconductors.''}

The intermediate state in type-I superconductors is irrelevant to the hydrides claimed to be high temperature type II superconductors
with strong pinning centers \cite{etrappedp}.
 
 In conclusion, we argue that the Comment \cite{talantsev_comment} has no merit. The points that we made in our paper Ref. \cite{nonstandard}
 are valid and important, an early warning that the resistance drops observed for hydrides under pressure may not be due to
 superconductivity. Since we wrote Ref. \cite{nonstandard}, substantial additional evidence has emerged that
 the hydrides are either ``nonstandard superconductors'' or, more likely, not superconductors \cite{nsr,hysteresis,hmtrapped2}.
 In addition, we point out that two papers \cite{csh,luh} claiming high temperature superconductivity in hydrides that showed anomalous behavior   in the resistance data \cite{hmnature,enormous} 
have been retracted \cite{cshretr,luhretr}  and a third one \cite{yttrium}
 is under investigation, and other claims \cite{hemleycsh,hemleyluh,hemley550}  have not been reproduced.\\
 
\begin{acknowledgments}
FM 
was supported in part by the Natural Sciences and Engineering
Research Council of Canada (NSERC).

\end{acknowledgments}


\begin{references}
 
\bibitem{nonstandard}
J. E. Hirsch and F. Marsiglio, ``Nonstandard superconductivity or no superconductivity in hydrides under high pressure'', \href{https://journals.aps.org/prb/abstract/10.1103/PhysRevB.103.134505}{Phys. Rev. B 103, 134505 (2021)}.
 
 \bibitem{talantsev_comment} E.F. Talantsev, V.S Minkov, F.F. Balakirev and M.I. Eremets, ``Comment on `Nonstandard Superconductivity or No Superconductivity in Hydrides under High
Pressure,' '' preceding comment,
\href{https://arxiv.org/abs/2311.07865}{arXiv:2311.07865 (2023)}.

        \bibitem{e2021p} V. S. Minkov,
 S. L. Budko, F. F. Balakirev , V. B. Prakapenka, S. Chariton, R. J. Husband,
H. P. Liermann and M. I. Eremets,, ``Magnetic field screening in hydrogen-rich high-temperature superconductors'',
\href{https://www.nature.com/articles/s41467-022-30782-x} {Nat Commun 13, 3194 (2022)}.

\bibitem{etrappedp}     V. S.  Minkov, V. Ksenofontov, S. L. Bud'ko, E. F. Talantsev and 
M. I. Eremets,
``Magnetic flux trapping in hydrogen-rich high-temperature superconductors'',
\href{https://www.nature.com/articles/s41567-023-02089-1}{Nat. Phys. (2023)}.

\bibitem{mgb2thin} N. Acharya, M. A. Wolak, T. Tan, N. Lee, A. C. Lang, M. Taheri, D. Cunnane, B. S. Karasik, and X. X. Xi,
``$MgB_2$  ultrathin films fabricated by hybrid physical chemical
vapor deposition and ion milling'', \href{https://pubs.aip.org/aip/apm/article/4/8/086114/121467/MgB2-ultrathin-films-fabricated-by-hybrid-physical}
{APL Mater. 4, 086114 (2016)}.
 
 
 \bibitem{wang2010} C.C. Wang, R. Zeng, X. Xu, S.X. Dou, ``Superconducting transition width under magnetic field in
MgB$_2$ polycrystalline samples'' \href{https://pubs.aip.org/aip/jap/article/108/9/093907/345551/Superconducting-transition-width-under-magnetic}{J. Appl. Phys 108, 093907 (2010)}.
 
 \bibitem{choi2017} W. J. Choi, Y. I. Seo, D. Ahmad and Yong Seung Kwon, ``Thermal activation energy of 3D vortex matter in NaFe$_{1?x}$Co$_x$As (x = 0.01, 0.03 and 0.07) single
crystals'' \href{https://www.nature.com/articles/s41598-017-11371-1}{Scientific Reports 7, 10900 (2017).}
 
 \bibitem{guo2020} Jing Guo, Yazhou Zhou, Cheng Huang, Shu Cai, Yutao Sheng, Genda Gu, Chongli Yang,
Gongchang Lin, Ke Yang, Aiguo Li, Qi Wu, Tao Xiang and Liling Sun, ``Crossover from two-dimensional to three-dimensional superconducting states in bismuth-based cuprate 
 superconductor,'' \href{https://www.nature.com/articles/s41567-019-0740-0}{Nature Physics 16, 295 (2020).}
 
 \bibitem{mozaffari2019} Shirin Mozaffari, Dan Sun, Vasily S. Minkov, Alexander P. Drozdov, Dmitry Knyazev, Jonathan B. Betts, Mari Einaga, Katsuya Shimizu, Mikhail I. Eremets, Luis Balicas andFedor F. Balakirev, ``Superconducting phase diagram of H$_3$S under high magnetic fields,''   
 \href{https://www.nature.com/articles/s41467-019-10552-y}{Nature Communications 10, 2522 (2019).}
 
 \bibitem{troyan2021}  Ivan A. Troyan, Dmitrii V. Semenok, Alexander G. Kvashnin, Andrey V. Sadakov, Oleg A. Sobolevskiy, Vladimir M. Pudalov, Anna G. Ivanova, Vitali B. Prakapenka, Eran Greenberg, Alexander G. Gavriliuk, Igor S. Lyubutin, Viktor V. Struzhkin, Aitor Bergara, Ion Errea, Raffaello Bianco, Matteo Calandra, Francesco Mauri, Lorenzo Monacelli, Ryosuke Akashi, Artem R. Oganov, ``Anomalous High-Temperature Superconductivity in YH6,'' \href{https://onlinelibrary.wiley.com/doi/abs/10.1002/adma.202006832}{Advanced Materials 33, 2006832 (2021).}
 
 \bibitem{troyan2023} Ivan A. Troyan, Dmitrii V. Semenok, Anna G. Ivanova, Andrey V. Sadakov, Di Zhou,
Alexander G. Kvashnin, Ivan A. Kruglov, Oleg A. Sobolevskiy, Marianna V. Lyubutina,
Dmitry S. Perekalin, Toni Helm, Stanley W. Tozer, Maxim Bykov, Alexander F. Goncharov,
Vladimir M. Pudalov, and Igor S. Lyubutin, ``Non-Fermi-Liquid Behavior of Superconducting SnH$_4$,''  \href{https://www.ncbi.nlm.nih.gov/pmc/articles/PMC10602579/}{Advanced Science 10, 2303622 (2023).}

\bibitem{recent} Jianning Guo, Dmitrii Semenok, Grigoriy Shutov, Di Zhou, Su Chen, Yulong Wang, Kexin Zhang, Xinyue Wu, Sven Luther, 
Toni Helm, Xiaoli Huang, Tian Cui,
``Unusual metallic state in superconducting A15-type $La4H_{23}$'',
\href{https://academic.oup.com/nsr/advance-article/doi/10.1093/nsr/nwae149/7651283}{National Science Review, nwae149 (2024)}.


\bibitem{blatter} G. Blatter, M. V. Feigel'man, V. B. Geshkenbein, A. I. Larkin, and V. M. Vinokur,
``Vortices in high-temperature superconductors'',
\href{https://journals.aps.org/rmp/abstract/10.1103/RevModPhys.66.1125}{Rev. Mod. Phys. 66, 1125 (1994)}.

\bibitem{hmnature} J. E. Hirsch and F. Marsiglio, ``Unusual width of the superconducting transition in a hydride'',
\href{https://www.nature.com/articles/s41586-021-03595-z}{Nature 596, E9  (2021)}.

  \bibitem{dc} M. Dogan and M.  L. Cohen, ``Anomalous behavior in high-pressure carbonaceous sulfur hydride'', 
\href{https://www.sciencedirect.com/science/article/pii/S0921453421000344}{Physica C 583, 1353851 (2021)}.

 \bibitem{enormous}
        J. E. Hirsch,
       ``Enormous variation in homogeneity and other anomalous features of room temperature superconductor samples'',
       \href{https://link.springer.com/article/10.1007/s10948-023-06593-6}{ J Supercond Nov Magn 36, 1489 (2023)}.

\bibitem{vdp}  J. E. Hirsch, ``Electrical resistance of hydrides under high pressure: evidence of superconductivity or confirmation bias?'', \href{https://link.springer.com/article/10.1007/s10948-023-06594-5}{  J Supercond Nov Magn 36, 1495 (2023)}.
 

 

\bibitem{nsr} J. E. Hirsch, ``Are hydrides under high-pressure–high-temperature superconductors?'',
\href{https://academic.oup.com/nsr/advance-article/doi/10.1093/nsr/nwad174/7202350}{National Science Review, nwad174 (2023} and
references therein.

  \bibitem{hysteresis} J. E. Hirsch, 
"Hysteresis loops in measurements of the magnetic moment of hydrides under high pressure: Implications for superconductivity'',
\href{https://www.sciencedirect.com/science/article/pii/S0921453424000145}
{Physica C 617,  1354449 (2024)}.

 \bibitem{hmtrapped2} J. E. Hirsch and F. Marsiglio, ``Further analysis of flux trapping experiments on hydrides under high pressure'',
 \href{https://www.sciencedirect.com/science/article/pii/S0921453424000650}{Physica C  620, 1 1354500 (2024)}.




       
             \bibitem{csh}  Elliot Snider, Nathan Dasenbrock-Gammon, Raymond McBride, Mathew Debessai, Hiranya Vindana, Kevin Vencatasamy, 
             Keith V. Lawler, Ashkan Salamat and Ranga P. Dias
     ``Room-temperature superconductivity in a carbonaceous sulfur hydride'',
     \href{https://www.nature.com/articles/s41586-020-2801-z}{Nature 586, 373 (2020)}.
     

     \bibitem{luh}
Nathan Dasenbrock-Gammon, Elliot Snider, Raymond McBride, Hiranya Pasan, Dylan Durkee, Nugzari Khalvashi-Sutter, Sasanka Munasinghe, Sachith E. Dissanayake, Keith V. Lawler, Ashkan Salamat and Ranga P. Dias, 
``Evidence of near-ambient superconductivity in a N-doped lutetium hydride'', 
     \href{https://www.nature.com/articles/s41586-023-05742-0}{Nature  615,  244 (2023)}.
     
                  \bibitem{cshretr}  Elliot Snider, Nathan Dasenbrock-Gammon, Raymond McBride, Mathew Debessai, Hiranya Vindana, Kevin Vencatasamy, Keith V. Lawler, Ashkan Salamat and Ranga P. Dias
     ``Retraction Note: Room-temperature superconductivity in a carbonaceous sulfur hydride'',
     \href{https://www.nature.com/articles/s41586-022-05294-9}{Nature 610, 804 (2022)}.
     

     \bibitem{luhretr}
Nathan Dasenbrock-Gammon, Elliot Snider, Raymond McBride, Hiranya Pasan, Dylan Durkee, Nugzari Khalvashi-Sutter, Sasanka Munasinghe, Sachith E. Dissanayake, Keith V. Lawler, Ashkan Salamat and Ranga P. Dias, 
``Retraction Note: Evidence of near-ambient superconductivity in a N-doped lutetium hydride'', 
   \href{https://www.nature.com/articles/s41586-023-06774-2}{Nature 624, 460 (2023)}.
     
     
     
           \bibitem{yttrium}  Elliot Snider, Nathan Dasenbrock-Gammon, Raymond McBride, Xiaoyu Wang, Noah Meyers, Keith V. Lawler, Eva Zurek, Ashkan Salamat, and Ranga P. Dias,
     ``Synthesis of Yttrium Superhydride Superconductor with a Transition Temperature up to 262 K by Catalytic Hydrogenation at High Pressures'',
     \href{https://journals.aps.org/prl/abstract/10.1103/PhysRevLett.126.117003}{Phys. Rev. Lett. 126, 117003 (2021)}.
     
      \bibitem{hemleycsh} Hiranya Pasan, Elliot Snider, Sasanka Munasinghe, Sachith E. Dissanayake, Nilesh P. Salke, Muhtar Ahart, Nugzari Khalvashi-Sutter, Nathan Dasenbrock-Gammon, Raymond McBride, G. Alexander Smith, Faraz Mostafaeipour, Dean Smith, Sergio Villa Cort\'es, Yuming Xiao, Curtis Kenney-Benson, Changyong Park, Vitali Prakapenka, Stella Chariton, Keith V. Lawler, Maddury Somayazulu, Zhenxian Liu, Russell J. Hemley, Ashkan Salamat, Ranga P. Dias,
     ``Observation of Conventional Near Room Temperature Superconductivity in Carbonaceous Sulfur Hydride'',
     \href{https://arxiv.org/abs/2302.08622}{arXiv:2302.08622 (2022)}.
     
     \bibitem{hemleyluh} Nilesh P. Salke, Alexander C. Mark, Muhtar Ahart and  Russell J. Hemley,
     ``Evidence for Near Ambient Superconductivity in the Lu-N-H System'',
     \href{https://arxiv.org/abs/2306.06301}{arXiv:2306.06301
 (2023)}.
     
     \bibitem{hemley550} A. D. Grockowiak, M. Ahart, T. Helm, W. A. Coniglio, R. Kumar, K. Glazyrin, G. Garbarino, Y. Meng, M. Oliff, V. Williams, N. W. Ashcroft,
R. J. Hemley, M. Somayazulu and S. W. Tozer,
     ``Hot Hydride Superconductivity Above 550 K'',
     \href{https://www.frontiersin.org/articles/10.3389/femat.2022.837651/full}{Front. Electron. Mater, 
Sec. Superconducting Materials
Volume 2 - 2022 | https://doi.org/10.3389/femat.2022.837651}.
 
   \end{references}
 \end{document}